\documentstyle[12pt]{article}
\newcommand{\be}{\begin{eqnarray}}
\newcommand{\ee}{\end{eqnarray}}

\newcommand{\yourtitle}[1]{
\mbox{}\\
\vskip 4\baselineskip
{\bf\noindent #1}\\ }

\newcommand{\youraddress}[1]{
\noindent\mbox{}\hspace{1in}\parbox[t]{4.5in}{#1}\\ }

\newcommand{\yournames}[1]{
\mbox{}\\
\mbox{}\\
\noindent\mbox{}\hspace{1in}{#1}\\ }

\newcommand{\yourabstract}[1]{
\mbox{}\\
\mbox{}\\
{\bf\noindent Abstract}\\
\begin{center}
\mbox{}\parbox[t]{5.in}{#1}
\end{center} }

\newcommand{\yoursection}[1]{
\vskip 2\baselineskip
{\bf\noindent #1}\\
\mbox{}\\
\vspace{-0.19in}}

\begin{document}

\yourtitle{HEAVY QUARK PRODUCTION IN $pp$ COLLISIONS}
\yournames{R. V. Gavai and S. Gupta}
\youraddress{Tata Institute, Bombay, India}
\yournames{P.L. McGaughey}
\youraddress{Los Alamos National Laboratory, Los Alamos, New Mexico}
\yournames{E. Quack}
\youraddress{Theory Department, \\
             Gesellschaft f\"ur Schwerionenforschung (GSI), \\
             Darmstadt, Germany}
\yournames{P. V. Ruuskanen\footnote{Department of Physics,
           University of Jyv\"askyl\"a, Jyv\"askyl\"a, Finland}}
\youraddress{Research Institute for Theoretical Physics,\\
            University of Helsinki, Helsinki, Finland}
\yournames{R. Vogt$^{\scriptsize \dag}$ and Xin-Nian Wang\footnote{Supported by
the U. S.
           Department of Energy under Contract No. DE-AC03-76SF00515.}}
\youraddress{Nuclear Science Division, MS 70A-3307\\
        Lawrence Berkeley Laboratory, \\
        University of California, Berkeley, California 94720}
\yourabstract{A systematic study of the inclusive single heavy
quark and heavy-quark pair
production cross sections in
$pp$ collisions is presented  for RHIC and LHC energies.  We compare with
existing data when possible.
The dependence  of  the rates on the renormalization
and factorization scales is discussed. Predictions
of the cross sections are given for two different sets of parton distribution
functions. }

\yoursection{INTRODUCTION}

%eq some stilistic changes
%Heavy quark production from initial nucleon-nucleon collisions will be copious
%at the RHIC and LHC colliders.  Charm and bottom production will represent a

Charm and bottom quark production from initial nucleon-nucleon collisions
will be copious at the RHIC and LHC colliders. Heavy quark decay into leptons
will represent a
significant background to dilepton production \cite{VJMR} in heavy ion
collisions.
% Since additional heavy quarks may also be produced by rescattering
% and in the quark-gluon plasma, it is important to have a good estimate of the
% production cross section in $pp$ collisions, calculable in perturbative QCD.
% eq instead of the above 3 lines:
A quantitative knowledge of the production cross section in $pp$ collisions
is a prerequisite for the detection of collective effects, such as heavy quark
production by rescattering and in the quark-gluon plasma, which appears as a
deviation from the simple superposition of hadronic collisions.

The lowest order (Born) calculations of the total cross section predict the
correct energy dependence but differ from the experimental measurements by a
``$K$ factor" of 2-3.   While the single-inclusive distributions as well as
the mass and rapidity distributions of $Q \overline{Q}$ pairs are also well
described to within a $K$ factor by the Born cross section, the $p_T$ and
azimuthal double-differential distributions are not calculable at the Born
level since the $Q \overline{Q}$ pair is always produced back-to-back
% eq
in lowest order. For
this reason, a next-to-leading order (NLO) calculation is needed.  The
calculations
%eq
we present here
are
done using a Monte Carlo
program developed by Nason and collaborators \cite{NDE1,NDE2,MNR}.
Similar work on the total cross section and the single inclusive distributions
was done by Smith, van Neerven, and collaborators \cite{SvN}.

%eq
In this calculation,
in addition to the uncertainties in the parton
distribution functions, uncertainties arise from the heavy quark mass and the
renormalization and factorization scale parameters.  At collider energies, the
calculations become more uncertain due to the lightness of the heavy quark
compared to the center of mass energy, $m_Q/\sqrt{s} \ll 1$.  We first
discuss the Born calculation in some detail and then outline the NLO
calculation with its additional uncertainties.  We use the
available data on
%eq $\sigma_{c \overline{c}}^{\rm tot}$
$\sigma_{c \overline{c}}^{\rm tot}(s)$
to fix the charm quark
mass and the scale parameters.  The resulting parameter set provides a point
from which to extrapolate to heavy-ion collider energies.
We then compare with single-inclusive
and double-differential distributions from charm and bottom data
when available.
We present estimates of heavy quark production cross sections in
proton-proton collisions at RHIC ($\sqrt{s} = 200$ and 500 GeV) and LHC
($\sqrt{s} = 5.5$ TeV and 14 TeV),
according to our present theoretical knowledge.  We provide both the Born
and NLO results for the total $Q \overline{Q}$ production cross section, single
inclusive $y$ and $p_{T}$ distributions, and double differential
$M$, $\phi$, $y$ and $p_T$ distributions.

\yoursection{HEAVY QUARK PRODUCTION IN PERTURBATIVE QCD}

The most general expression for the double differential cross section
for $Q \overline{Q}$ pair production
%eq is
from the collision of hadrons $A$ and $B$ is
\be
E_Q E_{\overline{Q}} \frac{d\sigma_{AB}}{d^3p_Q d^3p_{\overline{Q}}}
= \sum_{i,j}\int \,dx_1\, dx_2 F_i^A(x_1,\mu_F)F_j^B(x_2,\mu_F)
E_Q E_{\overline{Q}} \frac{d\widehat{\sigma}_{ij}
(x_1P_1,x_2P_2,m_Q,\mu_R)}{d^3p_Q d^3p_{\overline{Q}}} \,
\,  . \ee
% The functions $F_i$ are the number densities of gluons, light quarks and
% antiquarks evaluated at momentum fraction $x$ and factorization scale
%%$\mu_F$.
% Here $A$ and $B$ represent the initial hadrons and $i$, $j$ are the
%%interacting
% partons.  The short-distance cross section,
%eq instead of above 4 lines
Here $A$ and $B$ represent the initial hadrons and $i$, $j$ are the interacting
partons, and the functions $F_i$ are the number densities of gluons, light
quarks and antiquarks evaluated at momentum fraction $x$ and factorization
scale $\mu_F$. The short-distance cross section,
$\widehat{\sigma}_{ij}$, is calculable as a perturbation series in
$\alpha_s(\mu_R)$ where the strong coupling constant is evaluated at
the renormalization scale $\mu_R$.  Both scales are of the order of the
heavy quark mass.  At leading order, $\mu_F = \mu_R = \mu$ where $\mu = 2m_c$
is commonly used.  The scale
%eq variations
dependence
will be discussed in more detail below.

\yoursection{Leading Order}

%rv put LO comment here
At leading order, ${\cal O}(\alpha_s^2)$,
$Q \overline{Q}$ production proceeds by two basic processes,
\be
 q + \bar{q}  & \rightarrow & Q + \overline{Q} \\
 g + g & \rightarrow  & Q + \bar{Q}  \, \, .
\ee
The invariant cross section for the process $A + B \rightarrow
H + \overline{H} $ where the $Q \overline{Q}$ pair has fragmented into hadrons
%rv put quark content here--I hate extra parentheses
$H(Q \overline{q})$ and $\overline{H}(\overline{Q} q)$
can be written as
\be E_H E_{\overline{H}} \frac{d
\sigma_{AB}}{d^3p_H d^3p_{\overline{H}}} & = & \int \frac{\hat{s}}{2 \pi} dx_1
dx_2 dz_Q dz_{\overline{Q}} C(x_1,x_2) \frac{E_H E_{\overline{H}}}{E_Q
E_{\overline{Q}}} \\ \nonumber
&   & \frac{D_{H/Q}(z_Q)}{z_Q^3}
\frac{D_{\overline{H}/\overline{Q}}(z_{\overline
Q})}{z_{\overline{Q}}^3}
\delta^4 (p_1 + p_2 - p_Q - p_{\overline{Q}}) \, \, , \ee where
$\sqrt{\hat{s}}$, the
parton-parton center of mass energy, is related to $\sqrt{s}$, the
hadron-hadron center of mass energy, by $\hat{s} = x_1 x_2 s$.
The intrinsic
transverse momenta of the incoming partons have been neglected.
The sum of the
leading order subprocess
cross sections convoluted with the parton number densities is contained in
$C(x_1,x_2)$ where
\be C(x_1,x_2) = \sum_q
[F_q^A(x_1)  F_{\overline{q}}^B(x_2) +  F_{\overline{q}}^A(x_1) F_q^B(x_2)]
\frac{d
\hat{\sigma}_{q \overline{q}}}{d \hat{t}} + F_g^A(x_1) F_g^B(x_2) \frac{d
\hat{\sigma}_{gg}}{d \hat{t}} \, \, . \ee
Only light quark flavors, those with $m < m_Q$, are included in the sum
over $q$.  The dependence on the scale $\mu_F$ has been suppressed here.

Fragmentation affects the charmed hadron distributions, not the total
$c \overline{c}$ production cross section.  The fragmentation
functions, $D_{H/Q}(z)$, describe the hadronization of the heavy quarks
where $z = |\vec{p_H}|/|\vec{p_Q}|$ is the fraction of the heavy quark momentum
carried by the final-state hadron.  The $D$ meson $x_F$
distribution is harder than the calculated charmed quark distribution in
hadron-hadron interactions.  Including a fragmentation function that
describes $D$ production in $e^+ e^-$ annihilation softens
the distribution due to energy lost to light $q \overline{q}$ pair production
\cite{VB}.  Event generators such as PYTHIA \cite{PYT},
based on the Lund string fragmentation model, harden the $D$ distribution.
In PYTHIA, the charmed quark is always at the endpoint of a string which
pulls the
charmed quark in the direction of a beam remnant so that the charmed hadron
can be produced at a larger momentum than the charmed quark.  Correlations of
the produced charmed hadron with the projectile valence quarks, not predicted
by perturbative QCD, have been measured.  Several possible explanations have
been suggested, see {\it i.e.}, \cite{VB,PYT,MNR2}.  This interesting high
$x_F$ regime will not be measurable at the RHIC and LHC colliders since the
center of mass energy is high and the rapidity coverage is mostly confined to
the central region.  (The PHENIX muon spectrometer at RHIC will have a larger
rapidity coverage, $1.5 \leq y \leq 2.5$ \cite{CDR}, but these effects will
probably be out of reach at the maximum collider energy.)

If we ignore fragmentation effects for the moment,
%However, since only a small window of $x_F$ is open at RHIC and LHC, we will
%assume $D_{H/Q}(z) = \delta(z-1)$ and present the charmed quark distributions.
%Thus,
after taking four-momentum conservation into account, we are left with
\be \frac{d \sigma}{dp_T^2 dy_Q dy_{\overline{Q}}} = x_1 x_2
C(x_1,x_2) \, \, , \ee
where $x_1$ and $x_2$ are
\be x_1 & = & \frac{\widehat{m}_Q}{\sqrt{s}} (e^{y_Q} + e^{y_{\overline{Q}}})
\, \, , \\ \nonumber x_2 & = & \frac{\widehat{m}_Q}{\sqrt{s}}
(e^{-y_Q} + e^{-y_{\overline{Q}}})
\, \, , \ee and
$\widehat{m}_Q = \sqrt{m_Q^2 + p_T^2}$.  At $y_Q = y_{\overline{Q}} = 0$,
$x_1 = x_2$.  The target fractions, $x_2$, decrease with rapidity while the
projectile fractions, $x_1$, increase.
The subprocess cross sections for $Q \overline{Q}$ production by
$q\overline{q}$ annihilation and $gg$ fusion to order ${\cal O}(\alpha_s^2)$,
expressed as a function of
$\widehat{m}_Q$, $y_Q$, and $y_{\overline{Q}}$ are \cite{Ellis}
\be
\frac{d \hat{\sigma}_{q \overline{q}}}{d \hat{t}} = \frac{\pi
\alpha_s^2}{9 \widehat{m}_Q^4} \frac{\cosh(y_Q - y_{\overline{Q}}) +
m_Q^2/\widehat{m}_Q^2}{(1+ \cosh(y_Q - y_{\overline{Q}}))^3} \, \, ,
\ee
\be  \frac{d \hat{\sigma}_{gg}}{d \hat{t}} = \frac{\pi \alpha_s^2}{96
\widehat{m}_Q^4} \frac{8 \cosh(y_Q - y_{\overline{Q}}) - 1}{(1+
\cosh(y_Q - y_{\overline{Q}}))^3} \left(
\cosh(y_Q - y_{\overline{Q}}) + \frac{2m_Q^2}{\widehat{m}_Q^2} -
\frac{2m_Q^4}{\widehat{m}_Q^4} \right) \, \, .
\ee

\yoursection{Next-to-Leading Order}

%rv put NLO comment here instead
We now discuss the NLO, ${\cal O}(\alpha_s^3)$, corrections
to the $Q \overline{Q}$ production cross
section.  At next-to-leading order, in addition to virtual corrections to these
diagrams, production by
\be
 q + \bar{q} & \rightarrow  & Q + \bar{Q}  + g \\
 g + g & \rightarrow & Q + \overline{Q} + g\\
 q (\bar{q}) + g & \rightarrow & Q + \overline{Q}  + (\bar{q}) q \, \, ,
\ee
must also be included.  The last process, quark-gluon scattering, is not
present at leading order.  The quark-gluon
% graphs have been interpreted at the Born level as the scattering of a
% heavy quark excited from the nucleon sea with a light quark or gluon and
% referred to as flavor excitation \cite{NDE1}.
%eq instead of 3 lines above
graphs can be interpreted at the Born level as the scattering of a
heavy quark excited from the nucleon sea with a light quark or gluon and
are referred to as flavor excitation \cite{NDE1}.
The total short distance cross section $\widehat{\sigma}_{ij}$ for a given
production process can be expressed generally as \be
\widehat{\sigma}_{ij}(\widehat{s},m_Q,\mu_R) = \frac{\alpha_s^2(\mu_R)}{m_Q^2}
f_{ij}(\rho,\mu_R^2/m_Q^2) \, \, ,
\ee
where $\rho = 4m_Q^2/\widehat{s}$.  The function $f_{ij}$ can be expanded
perturbatively as \be
f_{ij}(\rho,\mu_R^2/m_Q^2) = f^0_{ij}(\rho)
+ \frac{\alpha_s(\mu_R)}{4\pi} \left[f^1_{ij}(\rho) + \overline
f^1_{ij}(\rho)\ln(\mu_R^2/m_Q^2) \right] + {\cal O}(\alpha_s^2)
\, \, . \ee
The leading order part of the cross section is in the function $f_{ij}^0$.
In this case, $f_{qg}^0 = f_{\overline{q} g}^0 = f_{gq}^0 = f_{g
\overline{q}}^0
 = 0$.  Only $f_{gg}^0$ and $f^0_{q \overline{q}}$ contribute and can be
computed from the $\hat{t}$ integration of the cross sections given in (8) and
(9).
%Ramona New discussion here
The physical cross section
should be independent of the renormalization scale: the dependence in
eq.\ (14) introduces an unphysical parameter in the calculation.  If the
perturbative expansion is sufficient, {\it i.e.} if further higher-order
corrections are small, at some value of $\mu$ the physical
${\cal O}(\alpha_s^n)$ and ${\cal O}(\alpha_s^{n+1})$ cross sections should be
equal\footnote{The order of the expansion is represented by $n$.
For $Q \overline{Q}$
production, $n\geq 2$.  A calculation to order ${\cal O}(\alpha_s^n)$
introduces corrections at the order ${\cal O}(\alpha_s^{n+1})$.}.  If the $\mu$
dependence is strong, the perturbative expansion is untrustworthy and the
predictive power of the calculation is weak \cite{Ellis}.  The rather large
difference between the heavy-quark Born and NLO cross sections suggests that
further higher-order corrections are needed, particularly for charm and bottom
quarks which are rather ``light" when $\sqrt{s}$ is large.
Usually the renormalization scale in $\widehat{\sigma}_{ij}$ and the
factorization scale in the parton distribution functions are chosen to be
equal.
We follow this prescription in our calculations.

We have used two sets of recent parton distribution
functions\footnote{All available parton
distribution functions are contained in the package PDFLIB \cite{PDF},
available in the CERN library routines.}, GRV HO \cite{GRV} and
MRS D-$^\prime$
\cite{D0}.  The first begins with a low scale, $Q_{0,{\rm GRV}}^2 =
0.3$ GeV$^2$, and
valence-like parton distributions, therefore evolving very quickly with $Q^2$.
The second, with $Q_{0,{\rm MRS}}^2 = 5$ GeV$^2$, has
sea quark and gluon distributions
that grow as $\sim x^{-1/2}$ when $x \rightarrow 0$.  Both are compatible
with the recent deep-inelastic scattering data from
HERA \cite{HERA}. We also include estimates of the total cross section using
the MRS D$0^\prime$ \cite{D0} distributions.  This set assumes a constant
value for the sea and gluon
distributions at $Q_{0,{\rm MRS}}^2$ as $x \rightarrow 0$
and lies below the HERA data.
The GRV distributions assume $\overline{u} = \overline{d}$, a symmetric light
quark sea, and $x \overline{s} (x,Q_{0,{\rm GRV}}^2) = 0$,
increasing to give $2 \langle x
\rangle_{\overline{s}}/(\langle x \rangle_{\overline{u}} + \langle x
\rangle_{\overline{d}}) \simeq
0.53$ at $Q^2 = 10$ GeV$^2$ \cite{GRV}.  The MRS D sets allow $\overline{u} <
\overline{d}$ to account for measurements of the Gottfried sum rule and assume
$\overline{s} = (\overline{u} + \overline{d})/4$ at $Q_{0,{\rm MRS}}^2$
\cite{D0}.  Thus the
MRS distributions, arising from a global fit, provide a somewhat better
description of the deep-inelastic scattering data for $x > 0.01$ than the GRV
distributions \cite{GRV,D0}.

Since we compare
two extreme cases for the nucleon parton distributions as $x \rightarrow 0$,
MRS
D-$^\prime$ and GRV HO on one hand and MRS D0$^\prime$ on the other, our
results may be thought of as providing an upper and lower bound to the
$Q \overline{Q}$ cross section at heavy-ion collider energies for fixed mass
and
scale.  However, little data
% eq exists
exist
on the gluon distribution function at low $x$ so that it is poorly
known, particularly in the $x$ region accessible at RHIC and LHC, $x \approx
10^{-2}$ and $10^{-4}$ around $y=0$, respectively.  The low $x$ behavior
has a significant
effect on the shape of the gluon distribution at moderate values of $x$
in the energy range of Fig.\ 1.  Steeply rising gluon distributions at low
$x$ are compensated for by a corresponding depletion at moderate $x$.

Heavy quark production by gluon fusion dominates the $pp \rightarrow Q
\overline
Q X$ production cross section in the central region.  Thus we show the shape
of the gluon distributions of the three parton distribution sets
are shown in Fig.\
1(a) over the $x$ range of the previous $pp$ data, $0.01 < x< 1$.
To facilitate comparison, all three are
shown at $\mu = 2.4$ GeV.  The solid curve is the GRV HO distribution, the
dashed, MRS D0$^\prime$, and the dot-dashed, MRS D-$^\prime$.  The GRV
distribution at $\mu = 1.2$ GeV is also shown to demonstrate
the effect of the $Q^2$
evolution.  Since it has a smaller initial scale, the evolution with $\mu$ is
quite fast.  The D0$^\prime$ distribution can be seen to turn over and begin
to flatten as $x$ decreases.  However, for much of the range, it is above the
D-$^\prime$ distribution, reflected in a larger $\sigma_{c \overline{c}}^{\rm
tot}$, as shown in Fig.\ 3.  All three sets, evaluated in the $\overline{MS}$
scheme, have a similar value of $\Lambda_{\rm QCD}$.  In Fig.\ 1(b),
we show the running of the two loop value of $\alpha_s$,
\be \alpha_s(\mu,f) = \frac{1}{b_f \ln(\mu^2/\Lambda_f^2)} \left[ 1 -
\frac{b_f^\prime \ln \ln(\mu^2/\Lambda_f^2)}{b_f \ln(\mu^2/\Lambda_f^2)}
\right]
\, \, , \ee
where $b_f = (33-2f)/12\pi$, $b_f^\prime = (153-19f)/(2\pi(33-2f))$, $f$ is
the number of flavors, and $\Lambda_f$ is the value of $\Lambda_{\rm QCD}$
appropriate for the number of flavors.  In the calculation, the number of
flavors depends on the chosen quark mass.  For charm, $f=3$, and for beauty,
$f=4$.  At $\mu = m_Q$, $\alpha_s(m_Q,f) = \alpha_s(m_Q,f+1)$.  The running of
$\alpha_s$ is visible in the renormalization scale dependence, shown in Fig.\
2(e).  For the NLO $Q \overline{Q}$ production program, $\Lambda_f$ is chosen
by $m_Q$.  Note that $\Lambda_3 > \Lambda_4 > \Lambda_5$.  Additional
uncertainties may arise because the threshold $m_Q$ for a given
parton distribution set can differ from our fitted $m_Q$.

%Ramona Your first question was my mistake.  I have checked this more carefully
%now.  Refer back to a new discussion under equation 14 to find out why I
%say the Born and NLO should produce the same results
While it is often possible to use a general prescription like the
principle of minimal sensitivity (PMS) \cite{PMS} to find values
of $\mu_R$ and $\mu_F$ where the scale sensitivity is a minimum, the
heavy quark production cross section is very sensitive
to changes in $\mu$.  In Fig.\ 2 we show the
variation of the $c \overline{c}$ and $b \overline{b}$ production cross
sections
at RHIC (a), (c) and LHC (b), (d) ion energies.  The MRS
distributions exhibit an artificial stability for low $\mu$ because
%eq when
for
$\mu
< 2m_c \approx Q_{0,{\rm MRS}}$, the factorization scale is fixed at
$Q_{0,{\rm MRS}}$ and only $\mu_R$
varies.  We use the GRV HO parton distribution
functions so that we can show the uncertainty with $\mu=\mu_R=\mu_F$ at lower
values of $\mu$ since $\mu_F$ is not fixed until $\mu_F \approx 0.4m_c \approx
Q_{0,{\rm GRV}}$.
When $\mu/m_c \approx 0.2$, the cross section
%eq blows up
diverges
since
$(\mu/m_c)/\Lambda_{\rm QCD} \approx 1$.
In any case, such small
scales below 1 GeV, are excluded because a perturbative calculation is no
longer assumed to be valid.
As $\mu/m_c$ increases, the cross section
becomes more stable.
%eq The results
The behavior we find is
similar for RHIC and LHC energies.
The $b \overline{b}$
cross section shows a smaller variation with $\mu$, particularly at $\sqrt{s}
= 200$ GeV.  The variation resembles the running of $\alpha_s$ shown in Fig.\
1(b).  Indeed, this running is a major source of instability in the NLO $Q
\overline{Q}$ cross sections.
%Vesa New comment (actually the old again. Don't worry about this now.)
% I still have difficulties with the "... the Born and the NLO
% calculations are equal, ..." logic but the reason is probably that I have
% never read the papers on PMS. I should.
However, at $\sqrt{s} = 5.5$ TeV the variation with $\mu$ at the Born
level increases since the cross section becomes more uncertain as
$m_Q/\sqrt{s}$ decreases.  The NLO results show less variation at this energy.
There is no value of $\mu$ where the Born and the NLO calculations are equal,
suggesting that higher-order corrections are needed for $m_Q/\sqrt{s} \ll 1$.

We show the change of the $c \overline{c}$ cross section at
$\sqrt{s} = 200$ GeV induced by fixing $\mu_R = 2m_Q$ and changing $\mu_F$ in
Fig.\ 2(e) and fixing $\mu_F=2m_Q$ and varying $\mu_R$ in Fig.\ 2(f).
The running of the coupling constant is clearly shown in 2(e).
In 2(f), the increase with $\mu_F$ arises because at values of $\mu_F$ near
$Q_{0,{\rm GRV}}$ and low $x$, the sea quark and gluon
distributions show a valence-like behavior, decreasing as $x \rightarrow 0$, an
effect special to the GRV distributions \cite{GRV}.  The results are quite
different for the MRS distributions, especially for the equivalent of Fig.\
2(f).  There is not much
change in the cross section with $\mu_F$, particularly at the Born level,
since the parton distribution functions do not change below $Q_{0,{\rm MRS}}$.

\yoursection{CALCULATIONS OF $\sigma_{Q \overline{Q}}^{\rm tot}$}

Previous comparisons of the total charm production cross sections with
calculations \cite{Ban} at leading order suggested that a constant $K$
factor of $\sim 2$ was needed to reconcile the calculations with data when
using
$m_c = 1.5$ GeV, but not when $m_c = 1.2$ GeV was chosen.  Initial NLO
calculations seemed to suggest that the $K$ factor was no longer needed with
$m_c = 1.5$ GeV \cite{Altarelli}.  However, this result is very dependent
upon the chosen scale parameters and the parton distribution functions,
particularly
the shape of the gluon distribution.

\yoursection{Comparison With Current Data}

%Ramona I put the NLO comment here and went into more detail about exact and
%inexact factors of two (the correction for all xf in pp is exact, the
%single to pair D cross section is inexact).
%eq
% We compare our NLO
We compare our NLO
calculations with the available data
\cite{Reu,AMM,E6531,NA27,NA321} on the total $c \overline
c$ production cross section from $pp$ and $pA$ interactions in Fig.\ 3.
When a nuclear target has been used, the cross section per nucleon is given,
assuming an $A^\alpha$ dependence with $\alpha=1$,
%eq addition:
%rv change
supported by recent experimental studies of the $A$
dependence \cite{alv}.
We assume that
we can compare the $c \overline{c}$ production cross section directly with
charmed hadron measurements.  Often single charmed mesons,
denoted $D/\overline{D}$ to include all charge states, in the region
$x_F > 0$ are measured.
The $c \overline{c}$ production cross
section is symmetric around $x_F = 0$ in $pp$ interactions so that $\sigma_{c
\overline{c}}^{\rm tot} = 2 \sigma_{c \overline{c}} (x_F >0)$.
While the question of how the $c \overline{c}$ pair
hadronizes into $D \overline{D}$, $D \overline{\Lambda_c}$, 
$\Lambda_c \overline{D}$, $\Lambda_c \overline{\Lambda_c}$,
{\it etc.}\ remains open, some
assumptions must be made about how much of $\sigma_{c \overline{c}}^{\rm tot}$
%eq
% is missing since all channels are not measured.  If all single
is missing since not all channels are measured.  If all single
$D$ mesons are assumed to originate from $D \overline{D}$ pairs,
ignoring associated $\Lambda_c \overline{D}$ production, then by definition
$\sigma(D \overline{D}) = \sigma(D/\overline{D})/2$.  Thus the single $D$ cross
section for $x_F >0$ is equal to the $D \overline{D}$ pair cross section over
all $x_F$.  However, the contribution to the $c
\overline{c}$ total cross section from $D_s$ and $\Lambda_c$ production has
been
estimated to be $\sigma(D_s)/\sigma(D^0 + D^+) \simeq 0.2$ and
$\sigma(\Lambda_c)/\sigma(D^0 + D^+) \simeq 0.3$.  Thus to obtain the total
$c \overline{c}$ cross section from $\sigma(D \overline{D})$, $\sigma(D
\overline{D})$ should be multiplied
%eq
% by $\sim 1.5$ \cite{MLM1}.
by $\approx 1.5$ \cite{MLM1}. This is done in our data comparison.
The data exist in the
range $19 < \sqrt{s} \leq 63$ GeV, mostly from fixed target experiments.  Below
the ISR energies, $\sqrt{s} = $ 53-63 GeV, the total cross section is
primarily inferred
from single $D$ or $D \overline{D}$ measurements.
At the ISR, the pair production cross
section is obtained from lepton measurements, either $e \mu$ and electron pair
coincidence measurements or
a lepton trigger in coincidence with a reconstructed $D$ or $\Lambda_c$.
Rather large $c \overline{c}$ cross sections were inferred from the latter
analyses due to the assumed shape of the production cross sections:  flat
distributions in $x_F$ for the $\Lambda_c$ and $(1-x_F)^3$ for the $D$.
The ISR results must thus be taken with some care.

Modern parton distributions with
$\Lambda_{\rm QCD}$ fixed by fits to data cannot explain the energy
dependence of the total cross section
%eq %rv also
in the measured energy range when using $m_c = 1.5$ GeV and
$\mu_F=\mu_R=m_c$.  Since $m_c^2 <
Q_{0,{\rm MRS}}^2$ for the MRS distributions and the scale must
be chosen so that
$\mu^2 > Q_{0,{\rm MRS}}^2$ for the calculations to make sense,
we take $\mu=2m_c$ and vary $m_c$ for these distributions.   We find reasonable
agreement for $m_c=1.2$ GeV for the D-$^\prime$ and D0$^\prime$ distributions.
The results are shown in the solid and dashed curves in Fig.\ 3 respectively.
Since the GRV HO distributions have a much lower initial scale, $\mu$
can be fixed to the quark mass.  The dot-dashed curve is the GRV HO
distribution with $m_c = 1.3$ GeV and $\mu = m_c$.  All three curves give
an equivalent description of the data.  Our ``fits" to the low energy data are
to provide a reasonable point from which to extrapolate to higher energies.
It is important to remember that significant uncertainties still exist which
could change our estimates considerably when accounted for.
These relatively low values of $m_c$
effectively provide an upper bound on the charm production cross section
at high energies.  For comparison, we also show the cross
section with the GRV distributions and $\mu = m_c = 1.5$ GeV in the dotted
curve.  It lies a factor of 2-3 below the other calculations.  The smaller
value of $m_c$ is needed for the MRS distributions even with the larger scale
because parton distribution functions at lower values of $Q^2$ would decrease
at low $x$, as demonstrated by the GRV distributions \cite{GRV}.  Note that
such small choices of $m_c$ suggests that the bulk of the total cross section
comes from invariant masses less than $2m_D$.  In a recent
work \cite{MLM1}, the total cross section data was
%eq
found to be
in agreement with $m_c =
1.5$ GeV with some essential caveats:  the factorization scale was fixed at
$\mu_F \equiv 2m_c$ while $\mu_R$ was allowed to vary and an older set of
parton distribution functions with a range of fits with a different value of
$\Lambda_{\rm QCD}$ for each was used.
Decreasing $\mu_R$ with respect to $\mu_F$ and
increasing $\Lambda_{\rm QCD}$ both result in a significantly larger cross
section for a given $m_c$. We choose
%eq
here
to use the most up-to-date parton
distribution functions and to keep $\mu_F = \mu_R$, facilitating a more direct
extrapolation from the current data to the future collider results.

Since data on $c \overline{c}$ and $b \overline{b}$ production by pion beams
%eq is
are
also available at fixed target energies, in Fig.\ 4 we show this data with
the same parton distributions where $m_c$ and $\mu$ are fixed by the comparison
in Fig.\ 3.  The $c \overline{c}$ data \cite{Reu,E515,NA322,E769,E6532}
is based on the $x_F > 0$ single $D$
cross section.  However, the $\pi^- N$ $x_F$ distribution is asymmetric,
$\sigma/\sigma(x_F >0) \sim 1.6$ so that $\sigma(D \overline{D})$ is obtained
by
dividing by 2 to get the pair cross section and then multiplying by 1.6 to
account for the partial $x_F$ coverage.  The $b \overline{b}$ data, taken to be
over all $x_F$, are generally obtained from multi-muon studies
\cite{NA10,E6533,WA78,E672}.
The data, especially for $b \overline{b}$ production, are not as extensive and
have rather poor statistics.  Again, some of the data is from a nuclear target.
When a nuclear target has been used, the cross section per nucleon is given,
assuming an $A^1$ dependence.

The GRV HO pion distributions \cite{GRVpi}
are based on their
proton set so that the two distributions are compatible.  In Fig.\ 4(a), the
charm production cross section is calculated using the GRV proton and pion
distributions.  The solid curve shows the result with a nucleon
target, the averaged distributions for proton and neutron, while the dashed
curve is the result for a proton target alone.  The results are consistent
at $\sqrt{s} = 30$ GeV; at lower energies, the cross section
%eq with
on
a proton target is slightly larger than
%eq with the
on a
nucleon target.
The calculations using
the MRS distributions do not have the same consistency as those with GRV
because their pion distribution functions, SMRS P1 and P2 \cite{SMRS}, are
based on an older set of proton distributions than the current MRS
distributions used here.
The SMRS distributions use $\Lambda_4 = 190$ MeV while the
MRS distributions have fixed $\Lambda_4 = 230$ MeV.  In the calculations, we
fix $\Lambda_4$ to the current MRS value.  The dot-dashed curve shows the
MRS D-$^\prime$ distributions with the SMRS P2 pion distributions while the
dotted curve is with the P1 set.  Both are for a proton target.  The P1 set
has a steeper gluon distribution
than P2.  The two calculations begin to diverge as $\sqrt{s}$ increases
since the gluon fusion contribution is becoming dominant.
At low $\sqrt{s}$,
valence quark annihilation is important for $\pi^- p$ interactions.
Although the calculations and data are not in exact agreement, they are
close enough to assume that the same parameters are reasonable for both pion
and proton projectiles.  The comparison to the $b \overline{b}$ production
cross section is given in Fig.\ 4(b).  The data is very sparse.  We use
$m_b = 4.75$ GeV and $\mu = m_b$ for both sets of parton distributions.  The
solid curve is the GRV distribution, the dashed is the MRS D-$^\prime$ and
SMRS P1 result.  The
agreement is not unreasonable given the quality of the data
%eq addition
on the one hand and the theoretical uncertainties on the other.

\yoursection{Extrapolation To RHIC And LHC Energies}

The total $c \overline{c}$ and $b \overline{b}$
cross sections at the top ISR energy, $\sqrt{s}=$ 63 GeV, and the
proton and ion
beam energies at RHIC and LHC are given in Tables 1 and 2 respectively.  Both
the Born and NLO cross sections are given.  The theoretical $K$ factor,
$\sigma_{Q
\overline{Q}}^{\rm NLO}/\sigma_{Q \overline{Q}}^{\rm LO}$, tends to increase
with
energy and is rather large.
There is no {\it a priori} reason why it should remain constant,
rather the increase at collider energies would suggest that the perturbative
%eq
expansion
is becoming less reliable, as discussed below.  Note that even though the MRS
D-$^\prime$ and GRV HO distributions give an equally valid description of the
data at ISR energies and below, they differ at higher energies, partly from
the difference in $m_c$ and partly because of our scale difference.  The
MRS D-$^\prime$ distributions evolve faster since $\mu = 2m_c$ rather than
$\mu = m_c$ due to their chosen initial scale $Q_{0,{\rm MRS}}$,
resulting in a larger predicted cross section.  Less difference is seen between
the GRV and MRS D-$^\prime$ distributions for the $b \overline{b}$ cross
section
since the $m_b$ and $\mu$ are used
for both.  Note that for $b \overline{b}$ production at 14 TeV, the results
differ by 30\% while the MRS D-$^\prime$ NLO $c \overline{c}$ result is three
times larger than the GRV HO result at the same energy.
The D0$^\prime$ distributions give smaller cross sections at LHC
energies due to the different initial behavior at $x \rightarrow 0$.
We illustrate this effect using the Born contribution
to the production cross section at fixed $M$ and $y=0$,
approximated as
\be \frac{d\sigma}{dMdy}|_{y=0} \approx \frac{\alpha_s^2}{Ms} \left[
F_g(M/\sqrt{s}) \right]^2 \, \,  \ee
since gluon fusion is the dominant contribution to the Born cross section,
$x=M/\sqrt{s}$ at $y=0$,
and at fixed $M$, $\sigma_{gg}$ is proportional to $(\alpha^2_s/M^2) F_g^2$.
The gluon distribution at low $x$ and $\mu=Q_0$ may be approximated as $F_g(x)
=
f(x)/x^{1+\delta}$.  For a constant behavior at low $x$,
such as in the MRS D0$^\prime$ distribution, $\delta=0$ and the cross section
is independent of $\sqrt{s}$.  At the other extreme, the MRS D-$^\prime$
distribution assumes $\delta = 0.5$ at $Q_0$
so that the cross section grows as $s^\delta \sim \sqrt{s}$.

\yoursection{SINGLE AND DOUBLE DIFFERENTIAL DISTRIBUTIONS}

We now compare the NLO calculations with data on $Q$ and $Q \overline{Q}$
distributions.
In the presentation of the single inclusive and double differential
distributions, we follow the
prescription of Nason and collaborators \cite{NDE2,MNR} and take
$\mu_S = n\widehat{m}_Q$ for the single and
$\mu_D = n \sqrt{m_Q + (p_{T_Q}^2 + p_{T_{\overline{Q}}}^2)/2}$ for the double
differential distributions.  When using MRS distributions for charm production,
$n=2$.  For all other cases, $n=1$.
A word of caution is necessary when looking at our predictions for $Q \overline
Q$ pair distributions.  It is difficult to properly regularize the soft
and collinear divergences to obtain a finite cross section over all phase
space.  Soft divergences cancel between real and virtual corrections when
properly regularized.  The collinear divergences need to be regularized and
subtracted.  For single inclusive heavy quark production, this is possible
%eq
% because analytic integration over the partonic recoil variables can be
% performed and the singularities isolated.
because the integration over the partonic recoil variables can be
performed analytically and the singularities isolated.
In exclusive $Q \overline{Q}$
pair production, the cancellation is
performed within the numerical integration.
The price paid for this is often a negative cross section near the phase
space boundaries, particularly when $p_{T} \rightarrow 0$ for the pair
and $\phi \rightarrow \pi$ where $\phi$ is the difference in the azimuthal
angle between the heavy quark and antiquark in the plane transverse to the
beam axis.  A positive differential cross section for $p_{T}
\rightarrow 0$ can only be obtained by resumming the full series of leading
Sudakov logarithms corresponding to an arbitrary number of soft gluons.
This has not been done in the case of heavy quark production \cite{MNR}.
Thus when $m_Q/\sqrt{s} \ll 1$, fluctuations in the cross section due to
incomplete numerical cancellations can become very large, resulting in negative
components in the mass and rapidity distributions.  We have minimized the
fluctuations by maximizing the event sampling at low $p_T$ and increasing the
number of iterations \cite{MLM2}.

\yoursection{Comparison To Current Data}

First, we compare with the 800 GeV fixed target data of the LEBC-MPS
collaboration \cite{AMM} in Fig.\ 5.
They measured the $x_F$ and $p_T^2$ distributions
of single $D$ production.  The total cross section, $\sigma(D/ \overline{D}) =
48 \pm 11 \, \mu$b, corresponds to a $D \overline{D}$ production cross section
of 24$\pm$8 $\mu$b.  The solid
curves are the MRS D-$^\prime$ results, the dashed, the GRV HO calculations.
Data on correlated $D \overline{D}$ production is also available
at 800 GeV, from $p$Emulsion studies \cite{Emu}.  The event sample is rather
small, only 35 correlated pairs.
We compare the mass and $p_T^2$ of the pair
and the azimuthal difference between the pair in Fig.\ 6 with the calculated
NLO distributions.  Again the solid curve is MRS D-$^\prime$, the dashed, GRV
HO.   The Born invariant mass
distribution, given by the dashed curve, is parallel to the NLO
results shown in the solid curve.

%Ramona Some changes made here.
The $p \overline{p}$ data from UA 1, $\sqrt{s} = 630$ GeV, and CDF, $\sqrt{s} =
1.8$ TeV, include single $b$ quark $p_T$ distributions.  The measurements are
taken in the central region ($|y| < 1.5$ for UA 1 and $|y|<1$ for CDF) and are
integrated over $p_T$ above each $p_{T,{\rm min}}$.
The comparisons with the NLO
calculations are given in Fig.\ 7(a) for UA 1 \cite{UA1} and Fig.\ 7(b) for CDF
and D0 \cite{CDF2,DZ}. Reasonable agreement is found for both GRV HO and MRS
D-$^\prime$ for UA 1 with $\mu_S = \sqrt{m_b^2 + p_T^2}$. However, the results
from this same scale choice lie somewhat below the early CDF data where data on
$J/\psi$ production was used to determine the $B$ production cross
section\footnote{The inclusive decay, $B \rightarrow J/\psi X$, has a 1\%
branching ratio ($BR$) while the channel $B \rightarrow J/\psi K$ has an 0.1\%
branching ratio.}. As reported in Ref.\ \cite{CDF1}, the scale $\mu = \mu_S/4$
was needed for good agreement with the magnitude of the data when the older MRS
D0 distributions were used. More recent data using direct measurement of
inclusive $b \rightarrow J/\psi$ and $b \rightarrow \psi^\prime$ decays has
shown that the previous results overestimated $\psi$ production from $b$ decays
\cite{CDF2}.  Better agreement with theory is now found for $\mu = \mu_S$, as
shown in Fig.\ 7(b).  Again the GRV HO and MRS D-$^\prime$ distributions look
similar, differing primarily for $p_{T,{\rm min}} < 10$ GeV. This difference is
increased for the lower scale choice where $\mu_S/4 < Q_{0,{\rm MRS}}$ for
$p_{T,{\rm min}} < 7.5$ GeV, cutting off the evolution of the MRS distributions
below this $p_{T,{\rm min}}$.  The GRV calculations evolve over all $p_{T,{\rm
min}}$ since $\mu_S/4 > Q_{0,{\rm GRV}}$, hence the larger difference.

\yoursection{Extrapolation To RHIC And LHC Energies}

%Ramona I moved some of the text around and removed the part about the
%fluctuations.  They are statistical and hopefully will be reduced once the
%current run finishes.
We now show the predicted heavy quark distributions for
RHIC ($\sqrt{s}$ = 200 and 500 GeV) and LHC ($\sqrt{s}$ = 5.5 and 14 TeV)
using the MRS D-$^\prime$ and GRV HO distributions.
The results are shown in Figs.\ 8-23.  We use the same scales on the $y$-axes
for both sets of parton distributions as much as possible to facilitate
comparison.  In each figure we show the single
quark $p_{T}$ (a) and $y$ (b) distributions and the $p_T$ (c),
rapidity (d), invariant
mass (e), and azimuthal angle (f) distributions of the $Q \overline{Q}$ pair.
The Born (LO) results are also given in (b), (d), and (e).  All the
distributions
have been divided by the corresponding bin width.  The single and pair
$p_T$ distributions are also given with the rapidity cuts $y < |1|$ at the LHC
and $y < |0.35|$ at RHIC, corresponding to the
%eq addition
planned
acceptances of ALICE
\cite{ALICE} and the PHENIX central detector \cite{CDR}.  These $p_T$
distributions are also divided by the width of the rapidity interval.
In Tables 3-10 we give
the $y$-integrated single $p_T^2$ NLO and Born distributions, the pair $p_T^2$
distributions with the cut on rapidity, and the NLO and Born invariant mass
distributions for $c$ and $b$ production at each energy with the MRS
D-$^\prime$
partons.  Note that all distributions have a 2 GeV bin width and that neither
it nor the rapidity bin width
has been removed in the tables.  The statistical
uncertainties are less than 1\% at low $p_T^2$ and $M$, increasing to 5-6\%
in the tails.  The uncertainty increases slightly with energy.

The development of a rapidity plateau can be seen in both the single and pair
rapidity distributions as the energy increases.  This plateau is generally
broader for the single quarks than the pair since the pair mass enters into
the estimate of the maximum pair rapidity while the smaller quark transverse
mass gives the maximum single quark rapidity.  The plateau is broader for
the MRS D-$^\prime$ parton distributions.  In the charm
rapidity distributions with the MRS D-$^\prime$ partons at 14 TeV, the plateau
edge is artificial.  The set has a minimum $x$ of $10^{-5}$, reached at
$y \sim 2.8$ for a single quark and a somewhat larger $y$ for the pair.  The
GRV HO distributions have a minimum $x = 10^{-6}$,
corresponding to $y \sim 4.5$, off the scale of our graphs.  The average quark
and pair $p_T$ increases with energy.  For charmed quarks, $\langle p_T^2
\rangle$ is larger for the pair than for a single quark.  The opposite result
is seen for $b$ quarks.  The GRV distributions result in larger $\langle p_T^2
\rangle$ than the MRS distributions.  Near $p_T \rightarrow 0$, the MRS parton
distributions show a steeper slope than the GRV distributions.  As $p_T$
increases, the slopes become somewhat similar at RHIC energies.

In general, the LO mass and rapidity distributions are nearly
equivalent to the NLO results scaled by a theoretical $K$
factor independent of $M$ and $y$.
%Large statistical fluctuations
%in the Born cross sections exist at the LHC energies, due
%in part to the smallness of the ratio $m_Q/\sqrt{s}$ at these energies.
At LHC energies, the expansion parameter becomes $\alpha_s \log(s/m_Q^2)$, of
order 1 for $m_Q/\sqrt{s} \ll 1$, spoiling the convergence of the perturbative
expansion \cite{MLM1}.  This causes our predictions to be less reliable at
these
energies.
Note that using $\mu_S$ for the single inclusive distributions
and $\mu_D$ for the double
differential distributions leads to somewhat different values of the
integrated NLO cross sections than given in Tables 1 and 2, calculated with
$\mu = nm_Q$,
since the correction terms grow with $\mu$.
The effect is relatively small for
the Born results since the faster evolution of the parton distribution
functions is partly compensated by the decrease
%eq in
of
$\alpha_s$ with increasing $\mu$.

%Xin-Nian, add here!
We also compare to the leading order charm distributions
obtained from HIJING \cite{XN} for the ion collider energies, 200 GeV (Figs.\
8,9) and 5.5 TeV (Figs.\ 12,13).  HIJING uses the same mass and scale
parameters
and parton distribution functions as the other calculations.
Although only a Born level calculation of $Q \overline{Q}$ production,
HIJING includes the effect of multiple parton showers which simulates aspects
of higher-order production (NLO includes the effect of only one additional
parton).  The rapidity distributions, shown for $y>0$ only, closely resemble
the
NLO calculations.  However, the $p_T^2$ distributions, taken in the rapidity
interval $|y|<2$ for the single $c$ quark and the pair, are softer,
especially for
the $c \overline{c}$ pair.  (Again, the distributions are divided by the
rapidity bin width.)  The distributions are also not strongly peaked at
low $p_T$, as are the NLO calculations, due to initial state radiation.
HIJING also
includes fragmentation of the $c \overline{c}$ pair into
hadrons.  The calculated $\phi$ distributions are not as sharply
peaked at $\phi = \pi$ as the NLO results.  Note also that the $D \overline{D}$
pair $\phi$ distributions from HIJING are more isotropic than the original
$c \overline{c}$ pairs.

\yoursection{$Q \overline{Q}$ Decays To Lepton Pairs}

Since heavy quark decays are an important
%eq part of
contribution to
the dilepton continuum,
we show $c \overline{c}$ and $b \overline{b}$ decays into dileptons at RHIC
and LHC for the MRS D-$^\prime$ sets.  Because heavy quark decays are not
incorporated into our double-differential
calculation, the heavy quark pairs have been created from the final
distributions.  The heavy quark
decays to leptons are thus calculated using a Monte Carlo program based on data
from $D$ decays at SLAC \cite{Balt} and $B$ decays from CLEO \cite{CLEO}.
The inclusive branching ratio for $D$ meson decay into a lepton, averaged over
charged and
neutral $D$'s is $BR(D^0/D^+ \rightarrow l^+ X) \sim 12$\%.  The corresponding
branching ratio for $B$ mesons of unspecified charge is
$BR(B \rightarrow l^+ X) \sim 10.4$\% \cite{PDG}.
$B$ decays represent a special challenge since lepton
pairs of opposite sign can be produced from the decay of a single $B$ by
$B \rightarrow D l X$ followed by $D \rightarrow l X$.  Thus the $B$
decays can produce dileptons from the following:  a combination of leptons
from a single $B$,
two leptons from primary $B$ decays, two leptons from secondary decays, and
a primary lepton from one $B$ and a secondary lepton from the opposite sign
$\overline{B}$.  The measurement of Ref.\ \cite{CLEO} is assumed to be
for primary $B$ decays to leptons. The NLO pair distributions
$d\sigma/dM$ and $d\sigma/dy$ agree well with a $K$ factor times the Born
results.  Therefore the correlated
distributions, $d\sigma/dMdy$, are calculated at leading order
%eq added - please check!
%rv--true enough, I forgot, Pat anyway only really needs the shape, I put in
%the normalization
and multiplied by this K-factor,
while the $p_T^2$ and $\phi$ distributions,
unavailable at leading order, are taken from the NLO results.  The heavy quark
pair is specified according to the correlated distributions from the calculated
cross section.  The momentum vectors of the individual quarks are computed
in the pair rest frame, using the rapidity gap between the quarks.  Once the
quark four-momenta have been specified, the decays are calculated in the
quark rest frame, according to the measured lepton momentum distributions, and
then boosted back to the nucleon-nucleon center of mass, the lab frame for
RHIC and LHC.  Finally, the pair quantities, $M_{ll}$, $y_{ll}$, and
$p_{T,{ll}}$, are computed.

The average number of $Q \overline{Q}$ pairs, $N_{Q \overline{Q}}$, produced
in a central nuclear collision is
estimated by multiplying the cross section from Tables 1 and 2 by
%eq added
the nuclear thickness
$T_{AB}(0)$.
If $N_{Q \overline{Q}}<1$, only correlated production is important.
The number of correlated lepton
pairs can be estimated by multiplying the number of $Q \overline{Q}$ pairs
by the square of the meson, $H$, branching ratio
to leptons: $N_{Q \overline{Q}} BR^2(H/\overline{H} \rightarrow l^\pm X)$.
However, if $N_{Q \overline{Q}}>1$, dilepton production from uncorrelated
$Q \overline{Q}$ pairs should be accounted for as well.
Then two $Q \overline{Q}$ pairs are generated
from the production cross section and the $Q$ from one pair is
decayed with the $\overline{Q}$ from the other.
Thus for uncorrelated $Q \overline{Q}$ production, the average number of
lepton pairs
is approximately $N_{Q \overline{Q}} (N_{Q \overline{Q}} -1)
BR^2(H \overline{H} \rightarrow l^\pm X)$ when $N_{Q \overline{Q}} \gg 1$.
If $N_{Q \overline{Q}} \approx 1$, a distribution in $N_{Q \overline{Q}}$ must
be considered to calculate the uncorrelated pairs.
%Vesa  The sentences below look fine. Let me keep the marker here since I
% don't have the figures and the text on paper.
In the following figures, we show
the correlated dilepton cross section in $pp$ collisions,
$\sigma_{ll} = BR^2(H/\overline{H} \rightarrow l^\pm X) \sigma_{Q
\overline{Q}}$.
In Fig.\ 27, showing uncorrelated lepton pairs from $D \overline{D}$ decays at
the LHC, we give the uncorrelated distributions with the value of the
correlated cross section since $N_{Q \overline{Q}}<1$ in $pp$ collisions.
To find the correct scale in central $AB$ collisions, calculate
$N_{Q \overline{Q}}$ and then multiply the lepton pair cross section by
$T_{AB}(0) (N_{Q \overline{Q}}-1)$.

In Figs.\ 24-25, we show the mass (a),
rapidity (b), and $p_T$ (c) distributions for the lepton pairs from $D
\overline{D}$ and $B \overline{B}$ pairs respectively.
The average mass of the lepton pairs from $D \overline{D}$ decays at RHIC
ion energies is $\langle M_{ll} \rangle = 1.35$ GeV and the average lepton pair
$p_T$,
$\langle p_{T,{ll}} \rangle = 0.8$
GeV;  from $B \overline{B}$ decays, $\langle M_{ll} \rangle = 3.17$ GeV and
$\langle p_{T,{ll}} \rangle = 1.9$ GeV.  A like-sign subtraction should
eliminate most of the uncorrelated charm production at RHIC.

%Ramona The number of ccbar pairs in Au collisions has gone up to about 9
%with the lower quark mass.  This comes out to about 1 uncorrelated lepton
%pair, not too many, but I will ask Pat if he thinks it is important.  I
%showed him an earlier version of the text and will send this file to him to
%look at as well--to go the text carefully through as you suggest.
%Vesa New comment: The point is that if n_QQbar > 2 then the number of
%  uncorrelated pairs is larger than the number of correlated pairs. This
%  without any cuts which may reduce more the uncorrelated pairs.
%Vesa You seem to indicate that the uncorrelated pairs are not important at
% RHIC. I don't have the numbers here now but if remember right the average
% number of pairs at RHIC was like 5. Then the number of uncorrelated lepton
% pairs (being proportional to N(N-1)) would be 4 times the number of
% correlated pairs (being proportional to N). Now, I don't know how the
% acceptance cuts affect these but I would not think that they sweep out
% everything. Uncorrelated pairs also have typically larger mass values but
% again the acceptance cuts are important here. Did you talk about this with
% Pat. He certainly is the person who probably has these things in his
% fingertips and should be encouraged to go the text carefully through.
 At LHC ion energies, the $c \overline{c}$ production cross sections are large
enough for uncorrelated charm production to be substantial and difficult
to subtract in nuclear collisions.
The average mass of the lepton pairs from correlated $D \overline{D}$ decays
here is $\langle M_{ll} \rangle = 1.46$ GeV and the
$\langle p_{T,{ll}} \rangle = 0.82$ GeV.
When the pairs are assumed to be uncorrelated,
then $\langle M_{ll} \rangle = 2.73$ GeV and
$\langle p_{T,{ll}} \rangle = 1$ GeV.  The average dilepton mass from
uncorrelated $D \overline{D}$
pairs is larger since the
rapidity gap between uncorrelated $D$ and $\overline{D}$ mesons is larger
on average than between correlated $D \overline{D}$ pairs.
The $b \overline{b}$ cross section is still small enough at the LHC for
uncorrelated lepton pair production from $B$ meson decays to be small.
However, the
acceptance for these pairs will be larger than for charm decays since
high mass lepton pairs from heavy quark decays have a large rapidity gap.
When acceptance cuts are applied, at least one member of a lepton pair will
have a large enough rapidity to escape undetected so that high mass pairs from
heavy quark decays will have a strongly reduced acceptance.  This reduction
will occur at larger values of $M_{ll}$ for $B \overline{B}$ than $D \overline
D$ decays.
 From all $B \overline{B}$ decays, $\langle M_{ll} \rangle = 3.39$ GeV and
$\langle p_{T,{ll}} \rangle = 2$ GeV.
%Vesa What are the units in these figures? Good.
%Ramona: units will be added on the plots
In Figs.\ 26-28, we show the mass (a),
rapidity (b), and $p_T$ (c) distributions for the dilepton pairs from
correlated and uncorrelated $D
\overline{D}$ and correlated $B \overline{B}$ pairs respectively.

\yoursection{ SUMMARY}

In this overview, we have attempted to use the theoretical state of the art
to predict heavy quark production in $pp$ collisions at RHIC and LHC energies.
%eq However,
Although much progress has been made in the higher-order calculations of $Q
\overline{Q}$ production, this is not meant to be
the final word.  Fragmentation and decay effects need to be incorporated into
our next-to-leading order calculations.  More structure function data from
HERA,
combined with collider data on jets and prompt photons, will produce further
refined sets of parton distribution functions.  Theoretical progress may allow
resummation at low $p_T$ or produce estimates of next-to-next-to-leading order
corrections.  New scale fixing techniques may result in a reduction of scale
uncertainties. Thus, there is still room for improvement in these calculations.
%eq final words added
Though the agreement with lower energy data allows us to extrapolate these
results to RHIC and LHC energies, major uncertainties still exist, particularly
at LHC energies. However, given our mass and scale parameters,
the GRV HO and MRS D-$^\prime$ parton distribution functions provide a rough
upper and lower limit on the theoretical predictions. This might be useful
in particular for the design of detectors at these facilities.\\

\yoursection{Acknowledgments}

We gratefully acknowledge the help of M. L. Mangano and G. Ridolfi with their
program package and G. Schuler for discussions.

\newpage
\vspace{1.0in}
\begin{table}
\begin{tabular}{|c|c|c|c|c|c|c|} \hline
                & \multicolumn{2}{c|}{MRS D0$^\prime$} &
\multicolumn{2}{c|}{GRV
HO} & \multicolumn{2}{c|}{MRS D-$^\prime$} \\ \cline{2-7}
$\sqrt{s}$(GeV) & $\sigma_{c \overline{c}}^{\rm LO}$ ($\mu$b) &
$\sigma_{c \overline{c}}^{\rm NLO}$ ($\mu$b) & $\sigma_{c \overline
c}^{\rm LO}$ ($\mu$b) & $\sigma_{c \overline{c}}^{\rm NLO}$ ($\mu$b) &
$\sigma_{c \overline{c}}^{\rm LO}$ ($\mu$b)
& $\sigma_{c \overline{c}}^{\rm NLO}$ ($\mu$b) \\ \hline
63 & 31.87 & 75.21 & 30.41 & 72.09 & 26.88 & 64.97 \\ \hline
200 & 105 & 244.2 & 122.6 & 350.8 & 139.3 & 343.7 \\ \hline
500 & 194.8 & 494 & 291.6 & 959 & 449.4 & 1138 \\ \hline
5500 & 558.2 & 1694 & 1687 & 6742 & 7013 & 17680 \\ \hline
14000 & 742.4 & 2323 & 2962 & 12440 & 16450 & 41770 \\ \hline
\end{tabular}
\caption{Total $c \overline{c}$ production cross sections at collider
energies.}
\end{table}
\vspace{1.5in}
\begin{table}
\begin{tabular}{|c|c|c|c|c|c|c|} \hline
                & \multicolumn{2}{c|}{MRS D0$^\prime$} &
\multicolumn{2}{c|}{GRV
HO} & \multicolumn{2}{c|}{MRS D-$^\prime$} \\ \cline{2-7}
$\sqrt{s}$(GeV) & $\sigma_{b \overline{b}}^{\rm LO}$ ($\mu$b) &
$\sigma_{b \overline{b}}^{\rm NLO}$ ($\mu$b) & $\sigma_{b \overline
b}^{\rm LO}$ ($\mu$b) & $\sigma_{b \overline{b}}^{\rm NLO}$ ($\mu$b) &
$\sigma_{b \overline{b}}^{\rm LO}$ ($\mu$b)
& $\sigma_{b \overline{b}}^{\rm NLO}$ ($\mu$b) \\ \hline
63 & 0.0458 & 0.0884 & 0.0366 & 0.0684 & 0.0397 & 0.0746 \\ \hline
200 & 0.981 & 1.82 & 0.818 & 1.51 & 0.796 & 1.47 \\ \hline
500 & 4.075 & 8.048 & 4.276 & 8.251 & 3.847 & 7.597 \\ \hline
5500 & 40.85 & 112 & 88.84 & 202.9 & 98.8 & 224 \\ \hline
14000 & 78.46 & 233.9 & 222.9 & 538.4 & 296.8 & 687.5 \\ \hline
\end{tabular}
\caption{Total $b \overline{b}$ production cross sections at collider
energies.}
\end{table}

\clearpage

\newpage
\begin{table}
\begin{tabular}{|c|c|c||c|c||c|c|c|}
\multicolumn{8}{c}{{\bf $c \overline{c}$ Production $\sqrt{s}=200$ GeV}}
\\ \hline
\multicolumn{3}{c||}{$d\sigma_c/dp_T^2$ ($\mu$b/2 GeV$^2$)} &
\multicolumn{2}{c||}{$d\sigma_{c \overline{c}}/dp_T^2dy$ ($\mu$b/2 GeV$^2$)} &
\multicolumn{3}{c}{$d\sigma_{c \overline{c}}/dM$ ($\mu$b/2 GeV)} \\ \hline
$p_T^2$ (GeV$^2$) & NLO & LO & $p_T^2$ (GeV$^2$) & NLO & $M$ (GeV) & NLO & LO
\\ \hline
1  & 232.5 & 102.2 & 1 & 30.90 & & & \\
3  & 37.93 & 15.14 & 3 & 3.916 & 3 & 172.8 & 76.41 \\
5  & 12.37 & 4.589 & 5 & 1.548 & 5 & 77.05 & 34.18 \\
7  & 5.362 & 1.924 & 7 & 0.8435 & 7 & 22.60 & 9.611 \\
9  & 2.774 & 0.9704 & 9 & 0.4770 & 9 & 8.548 & 3.429 \\
11 & 1.589 & 0.5435 & 11 & 0.3287 & 11 & 3.671 & 1.427 \\
13 & 1.003 & 0.3389 & 13 & 0.2203 & 13 & 1.863 & 0.6871 \\
15 & 0.6715 & 0.2206 & 15 & 0.1608 & 15 & 0.9122 & 0.3438 \\
17 & 0.4612 & 0.1542 & 17 & 0.1277 & 17 & 0.5120 & 0.1917 \\
19 & 0.3291 & 0.1079 & 19 & 0.0925 & 19 & 0.3154 & 0.1095 \\
21 & 0.2399 & 0.0812 & 21 & 0.0786 & 21 & 0.1883 & 0.0651 \\
23 & 0.1857 & 0.0602 & 23 & 0.0589 & 23 & 0.1210 & 0.0415 \\
25 & 0.1369 & 0.0428 & 25 & 0.0478 & 25 & 0.0689 & 0.0245 \\
27 & 0.1088 & 0.0355 & 27 & 0.0356 & 27 & 0.0520 & 0.0166 \\
29 & 0.0864 & 0.0280 & 29 & 0.0350 & 29 & 0.0364 & 0.0105 \\
31 & 0.0697 & 0.0225 & 31 & 0.0282 & 31 & 0.0257 & 0.00785 \\
33 & 0.0574 & 0.0191 & 33 & 0.0206 & 33 & 0.0151 & 0.00538 \\
35 & 0.0478 & 0.0160 & 35 & 0.0214 & 35 & 0.0111 & 0.00383 \\
37 & 0.0400 & 0.0132 & 37 & 0.0160 & 37 & 0.0678 & 0.00222 \\
39 & 0.0343 & 0.0111 & 39 & 0.0135 & 39 & 0.0480 & 0.00198 \\ \hline
\end{tabular}
\caption{}[The rapidity-integrated $p_T^2$ distribution is given for
single charm (NLO and Born) and the $p_T^2$ distribution in the range
$|y|<0.35$ is given for $c \overline{c}$ pair production (NLO only).  The
tabulated results have not been corrected for the rapidity bin width.
The rapidity-integrated pair mass distribution is also given.  All
distributions
are at $\sqrt{s}=200$ GeV and calculated with MRS D-$^\prime$ parton
distributions.  Note the 2 GeV bin width for the distributions.]
\end{table}
\newpage

\begin{table}
\begin{tabular}{|c|c|c||c|c||c|c|c|}
\multicolumn{8}{c}{{\bf $c \overline{c}$ Production $\sqrt{s}=500$ GeV}}
\\ \hline
\multicolumn{3}{c||}{$d\sigma_c/dp_T^2$ ($\mu$b/2 GeV$^2$)} &
\multicolumn{2}{c||}{$d\sigma_{c \overline{c}}/dp_T^2dy$ ($\mu$b/2 GeV$^2$)} &
\multicolumn{3}{c}{$d\sigma_{c \overline{c}}/dM$ ($\mu$b/2 GeV)} \\ \hline
$p_T^2$ (GeV$^2$) & NLO & LO & $p_T^2$ (GeV$^2$) & NLO & $M$ (GeV) & NLO & LO
\\ \hline
1  & 739.7 & 332.0 & 1 & 68.64 & & & \\
3  & 134.8 & 538.7 & 3 & 12.01 & 3 & 548.1 & 242.7 \\
5  & 47.37 & 17.43 & 5 & 4.874 & 5 & 259.5 & 117.2 \\
7  & 22.19 & 7.656 & 7 & 2.828 & 7 & 82.67 & 35.73 \\
9  & 12.08 & 4.054 & 9 & 1.809 & 9 & 32.71 & 13.72 \\
11 & 7.336 & 2.400 & 11 & 1.193 & 11 & 15.19 & 6.223 \\
13 & 4.658 & 1.493 & 13 & 0.8440 & 13 & 7.878 & 3.108 \\
15 & 3.281 & 1.041 & 15 & 0.6417 & 15 & 4.623 & 1.734 \\
17 & 2.343 & 0.7234 & 17 & 0.5002 & 17 & 2.555 & 1.025 \\
19 & 1.758 & 0.5370 & 19 & 0.3983 & 19 & 1.577 & 0.6242 \\
21 & 1.328 & 0.3980 & 21 & 0.3345 & 21 & 1.143 & 0.4171 \\
23 & 1.034 & 0.3052 & 23 & 0.2467 & 23 & 0.7373 & 0.2623 \\
25 & 0.8118 & 0.2512 & 25 & 0.2098 & 25 & 0.4798 & 0.1905 \\
27 & 0.6481 & 0.1950 & 27 & 0.1596 & 27 & 0.3227 & 0.1220 \\
29 & 0.5411 & 0.1618 & 29 & 0.1371 & 29 & 0.2817 & 0.0886 \\
31 & 0.4544 & 0.1284 & 31 & 0.1283 & 31 & 0.2028 & 0.0673 \\
33 & 0.3600 & 0.0997 & 33 & 0.1137 & 33 & 0.1530 & 0.0472 \\
35 & 0.3006 & 0.0897 & 35 & 0.0909 & 35 & 0.0997 & 0.0379 \\
37 & 0.2701 & 0.0754 & 37 & 0.0758 & 37 & 0.0837 & 0.0293 \\
39 & 0.2318 & 0.0643 & 39 & 0.0750 & 39 & 0.0627 & 0.0250 \\ \hline
\end{tabular}
\caption{}[The rapidity-integrated $p_T^2$ distribution is given for
single charm (NLO and Born) and the $p_T^2$ distribution in the range
$|y|<0.35$ is given for $c \overline{c}$ pair production (NLO only).
The
tabulated results have not been corrected for the rapidity bin width.
The rapidity-integrated pair mass distribution is also given.  All
distributions
are at $\sqrt{s}=500$ GeV and calculated with MRS D-$^\prime$ parton
distributions.  Note the 2 GeV bin width for the distributions.]
\end{table}
\newpage

\begin{table}
\begin{tabular}{|c|c|c||c|c||c|c|c|}
\multicolumn{8}{c}{{\bf $c \overline{c}$ Production $\sqrt{s}=5.5$ TeV}}
\\ \hline
\multicolumn{3}{c||}{$d\sigma_c/dp_T^2$ ($\mu$b/2 GeV$^2$)} &
\multicolumn{2}{c||}{$d\sigma_{c \overline{c}}/dp_T^2dy$ ($\mu$b/2 GeV$^2$)} &
\multicolumn{3}{c}{$d\sigma_{c \overline{c}}/dM$ ($\mu$b/2 GeV)} \\ \hline
$p_T^2$ (GeV$^2$) & NLO & LO & $p_T^2$ (GeV$^2$) & NLO & $M$ (GeV) & NLO & LO
\\ \hline
1  & 10680. & 5146. & 1 & 1840. & & & \\
3  & 2453. & 989. & 3 & 441.5 & 3 & 7749. & 3558. \\
5  & 974.8 & 350.1 & 5 & 196.9 & 5 & 4366. & 2048. \\
7  & 502.2 & 166.9 & 7 & 111.3 & 7 & 1622. & 709.2 \\
9  & 289.8 & 93.10 & 9 & 75.68 & 9 & 693.7 & 297.5 \\
11 & 186.6 & 57.12 & 11 & 51.60 & 11 & 351. & 144.0 \\
13 & 126.4 & 37.65 & 13 & 39.07 & 13 & 188.9 & 78.77 \\
15 & 90.91 & 25.96 & 15 & 27.28 & 15 & 116.3 & 45.67 \\
17 & 68.95 & 19.99 & 17 & 22.55 & 17 & 75.79 & 27.83 \\
19 & 51.44 & 14.43 & 19 & 18.47 & 19 & 50.16 & 18.82 \\
21 & 41.11 & 11.17 & 21 & 14.14 & 21 & 30.89 & 12.54 \\
23 & 33.29 & 8.965 & 23 & 13.53 & 23 & 23.02 & 9.024 \\
25 & 27.23 & 7.328 & 25 & 11.02 & 25 & 18.04 & 6.489 \\
27 & 22.28 & 6.031 & 27 & 9.862 & 27 & 12.32 & 4.547 \\
29 & 18.64 & 4.836 & 29 & 8.612 & 29 & 10.75 & 3.635 \\
31 & 16.10 & 4.203 & 31 & 6.944 & 31 & 8.112 & 2.609 \\
33 & 13.51 & 3.417 & 33 & 6.359 & 33 & 5.596 & 2.038 \\
35 & 11.55 & 2.961 & 35 & 5.050 & 35 & 5.217 & 1.719 \\
37 & 9.881 & 2.548 & 37 & 4.683 & 37 & 4.214 & 1.240 \\
39 & 9.078 & 2.212 & 39 & 4.680 & 39 & 3.500 & 1.039 \\ \hline
\end{tabular}
\caption{}[The rapidity-integrated $p_T^2$ distribution is given for
single charm (NLO and Born) and the $p_T^2$ distribution in the range
$|y|<1$ is given for $c \overline{c}$ pair production (NLO only).
The
tabulated results have not been corrected for the rapidity bin width.
The rapidity-integrated pair mass distribution is also given.  All
distributions
are at $\sqrt{s}=5.5$ TeV and calculated with MRS D-$^\prime$ parton
distributions.  Note the 2 GeV bin width for the distributions.]
\end{table}
\newpage

\begin{table}
\begin{tabular}{|c|c|c||c|c||c|c|c|}
\multicolumn{8}{c}{{\bf $c \overline{c}$ Production $\sqrt{s}=14$ TeV}}
\\ \hline
\multicolumn{3}{c||}{$d\sigma_c/dp_T^2$ ($\mu$b/2 GeV$^2$)} &
\multicolumn{2}{c||}{$d\sigma_{c \overline{c}}/dp_T^2dy$ ($\mu$b/2 GeV$^2$)} &
\multicolumn{3}{c}{$d\sigma_{c \overline{c}}/dM$ ($\mu$b/2 GeV)} \\ \hline
$p_T^2$ (GeV$^2$) & NLO & LO & $p_T^2$ (GeV$^2$) & NLO & $M$ (GeV) & NLO & LO
\\ \hline
1  & 23650. & 11960. & 1 & 4594. & & & \\
3  & 6067. & 2473. & 3 & 1129. & 3 & 17250. & 8046. \\
5  & 2576. & 918.6 & 5 & 513.6 & 5 & 10240. & 4960. \\
7  & 1368. & 452.4 & 7 & 298.9 & 7 & 4119. & 1840. \\
9  & 838.8 & 256.5 & 9 & 195.3 & 9 & 1875. & 820.2 \\
11 & 545.2 & 162.7 & 11 & 143.4 & 11 & 986.3 & 413.9 \\
13 & 371.4 & 108.3 & 13 & 103.9 & 13 & 554.6 & 232.4 \\
15 & 273.5 & 78.46 & 15 & 78.28 & 15 & 337.7 & 137.8 \\
17 & 206.6 & 55.28 & 17 & 60.18 & 17 & 226.5 & 88.37 \\
19 & 162.1 & 45.82 & 19 & 51.11 & 19 & 162. & 57.77 \\
21 & 130.4 & 33.90 & 21 & 40.63 & 21 & 107.4 & 41.12 \\
23 & 102.5 & 26.90 & 23 & 34.76 & 23 & 71.90 & 28.14 \\
25 & 84.26 & 22.64 & 25 & 28.13 & 25 & 59.46 & 21.23 \\
27 & 70.85 & 18.27 & 27 & 24.60 & 27 & 38.62 & 15.25 \\
29 & 60.26 & 15.58 & 29 & 21.12 & 29 & 30.19 & 12.05 \\
31 & 51.43 & 13.08 & 31 & 17.05 & 31 & 25.45 & 8.619 \\
33 & 45.92 & 11.02 & 33 & 17.66 & 33 & 22.84 & 6.839 \\
35 & 40.26 & 9.718 & 35 & 16.21 & 35 & 15.55 & 5.642 \\
37 & 33.92 & 7.860 & 37 & 12.86 & 37 & 13.24 & 4.484 \\
39 & 29.80 & 7.281 & 39 & 10.61 & 39 & 11.64 & 3.454 \\ \hline
\end{tabular}
\caption{}[The rapidity-integrated $p_T^2$ distribution is given for
single charm (NLO and Born) and the $p_T^2$ distribution in the range
$|y|<1$ is given for $c \overline{c}$ pair production (NLO only).
The
tabulated results have not been corrected for the rapidity bin width.
The rapidity-integrated pair mass distribution is also given.  All
distributions
are at $\sqrt{s}=14$ TeV and calculated with MRS D-$^\prime$ parton
distributions.  Note the 2 GeV bin width for the distributions.]
\end{table}
\newpage

\begin{table}
\begin{tabular}{|c|c|c||c|c||c|c|c|}
\multicolumn{8}{c}{{\bf $b \overline{b}$ Production $\sqrt{s}=200$ GeV}}
\\ \hline
\multicolumn{3}{c||}{$d\sigma_b/dp_T^2$ ($\mu$b/2 GeV$^2$)} &
\multicolumn{2}{c||}{$d\sigma_{b \overline{b}}/dp_T^2dy$ ($\mu$b/2 GeV$^2$)} &
\multicolumn{3}{c}{$d\sigma_{b \overline{b}}/dM$ ($\mu$b/2 GeV)} \\ \hline
$p_T^2$ (GeV$^2$) & NLO & LO & $p_T^2$ (GeV$^2$) & NLO & $M$ (GeV) & NLO & LO
\\ \hline
1  & 0.2201 & 0.1123 & 1 & 0.2073 & & & \\
3  & 0.1704 & 0.0883 & 3 & 0.0524 & & & \\
5  & 0.1558 & 0.0680 & 5 & 0.0263 & & & \\
7  & 0.1064 & 0.0541 & 7 & 0.0170 & & & \\
9  & 0.1035 & 0.0577 & 9 & 0.0118 & 9 & 0.0463 & 0.0320 \\
11 & 0.0863 & 0.0406 & 11 & 0.00814 & 11 & 0.4363 & 0.2100 \\
13 & 0.0605 & 0.0343 & 13 & 0.00660 & 13 & 0.3184 & 0.1640 \\
15 & 0.0478 & 0.0255 & 15 & 0.00441 & 15 & 0.1987 & 0.1050 \\
17 & 0.0458 & 0.0264 & 17 & 0.00341 & 17 & 0.1225 & 0.0637 \\
19 & 0.0351 & 0.0190 & 19 & 0.00311 & 19 & 0.0753 & 0.0400 \\
21 & 0.0359 & 0.0186 & 21 & 0.00274 & 21 & 0.0492 & 0.0249 \\
23 & 0.0300 & 0.0139 & 23 & 0.00237 & 23 & 0.0318 & 0.0160 \\
25 & 0.0244 & 0.0122 & 25 & 0.00201 & 25 & 0.0214 & 0.0104 \\
27 & 0.0216 & 0.0116 & 27 & 0.00183 & 27 & 0.0145 & 0.00688 \\
29 & 0.0202 & 0.0103 & 29 & 0.00156 & 29 & 0.0091 & 0.00466 \\
31 & 0.0171 & 0.0080 & 31 & 0.00147 & 31 & 0.0069 & 0.00321 \\
33 & 0.0159 & 0.0083 & 33 & 0.00121 & 33 & 0.0047 & 0.00215 \\
35 & 0.0125 & 0.0054 & 35 & 0.00111 & 35 & 0.0032 & 0.00154 \\
37 & 0.0101 & 0.0055 & 37 & 0.00111 & 37 & 0.0022 & 0.00108 \\
39 & 0.0097 & 0.0049 & 39 & 0.00086 & 39 & 0.0016 & 0.00075 \\ \hline
\end{tabular}
\caption{}[The rapidity-integrated $p_T^2$ distribution is given for
single $b$ quarks(NLO and Born) and the $p_T^2$ distribution in the range
$|y|<0.35$ is given for $b \overline{b}$ pair production (NLO only).
The
tabulated results have not been corrected for the rapidity bin width.
The rapidity-integrated pair mass distribution is also given.  All
distributions
are at $\sqrt{s}=200$ GeV and calculated with MRS D-$^\prime$ parton
distributions.  Note the 2 GeV bin width for the distributions.]
\end{table}
\newpage

\begin{table}
\begin{tabular}{|c|c|c||c|c||c|c|c|}
\multicolumn{8}{c}{{\bf $b \overline{b}$ Production $\sqrt{s}=500$ GeV}}
\\ \hline
\multicolumn{3}{c||}{$d\sigma_b/dp_T^2$ ($\mu$b/2 GeV$^2$)} &
\multicolumn{2}{c||}{$d\sigma_{b \overline{b}}/dp_T^2dy$ ($\mu$b/2 GeV$^2$)} &
\multicolumn{3}{c}{$d\sigma_{b \overline{b}}/dM$ ($\mu$b/2 GeV)} \\ \hline
$p_T^2$ (GeV$^2$) & NLO & LO & $p_T^2$ (GeV$^2$) & NLO & $M$ (GeV) & NLO & LO
\\ \hline
1  & 0.9809 & 0.4798 & 1 & 0.3427 & & & \\
3  & 0.7911 & 0.4024 & 3 & 0.2503 & & & \\
5  & 0.6490 & 0.3362 & 5 & 0.1260 & & & \\
7  & 0.5492 & 0.2801 & 7 & 0.0818 & & & \\
9  & 0.4528 & 0.2358 & 9 & 0.0558 & 9 & 0.2652 & 0.1199 \\
11 & 0.3807 & 0.1987 & 11 & 0.0426 & 11 & 1.737 & 0.8711 \\
13 & 0.3256 & 0.1688 & 13 & 0.0341 & 13 & 1.436 & 0.7552 \\
15 & 0.2781 & 0.1433 & 15 & 0.0285 & 15 & 0.9909 & 0.5222 \\
17 & 0.2428 & 0.1248 & 17 & 0.0235 & 17 & 0.6646 & 0.3503 \\
19 & 0.2068 & 0.1057 & 19 & 0.0197 & 19 & 0.4547 & 0.2356 \\
21 & 0.1824 & 0.0932 & 21 & 0.0169 & 21 & 0.3132 & 0.1612 \\
23 & 0.1595 & 0.0811 & 23 & 0.0147 & 23 & 0.2183 & 0.1121 \\
25 & 0.1429 & 0.0719 & 25 & 0.0133 & 25 & 0.1566 & 0.0797 \\
27 & 0.1240 & 0.0622 & 27 & 0.0122 & 27 & 0.1126 & 0.0578 \\
29 & 0.1108 & 0.0557 & 29 & 0.0109 & 29 & 0.0850 & 0.0419 \\
31 & 0.0984 & 0.0492 & 31 & 0.0098 & 31 & 0.0640 & 0.0315 \\
33 & 0.0898 & 0.0435 & 33 & 0.0085 & 33 & 0.0469 & 0.0236 \\
35 & 0.0789 & 0.0387 & 35 & 0.0076 & 35 & 0.0367 & 0.0179 \\
37 & 0.0716 & 0.0350 & 37 & 0.0071 & 37 & 0.0291 & 0.0138 \\
39 & 0.0646 & 0.0319 & 39 & 0.0074 & 39 & 0.0220 & 0.0108 \\ \hline
\end{tabular}
\caption{}[The rapidity-integrated $p_T^2$ distribution is given for
single $b$ quarks(NLO and Born) and the $p_T^2$ distribution in the range
$|y|<0.35$ is given for $b \overline{b}$ pair production (NLO only).
The
tabulated results have not been corrected for the rapidity bin width.
The rapidity-integrated pair mass distribution is also given.  All
distributions
are at $\sqrt{s}=500$ GeV and calculated with MRS D-$^\prime$ parton
distributions.  Note the 2 GeV bin width for the distributions.]
\end{table}
\newpage

\begin{table}
\begin{tabular}{|c|c|c||c|c||c|c|c|}
\multicolumn{8}{c}{{\bf $b \overline{b}$ Production $\sqrt{s}=5.5$ TeV}}
\\ \hline
\multicolumn{3}{c||}{$d\sigma_b/dp_T^2$ ($\mu$b/2 GeV$^2$)} &
\multicolumn{2}{c||}{$d\sigma_{b \overline{b}}/dp_T^2dy$ ($\mu$b/2 GeV$^2$)} &
\multicolumn{3}{c}{$d\sigma_{b \overline{b}}/dM$ ($\mu$b/2 GeV)} \\ \hline
$p_T^2$ (GeV$^2$) & NLO & LO & $p_T^2$ (GeV$^2$) & NLO & $M$ (GeV) & NLO & LO
\\ \hline
1  & 23.59 & 11.22 & 1 & -2.366 & & & \\
3  & 19.38 & 9.650 & 3 & 12.80 & & & \\
5  & 16.25 & 8.253 & 5 & 6.634 & & & \\
7  & 13.84 & 7.028 & 7 & 4.424 & & & \\
9  & 11.83 & 6.065 & 9 & 3.303 & 9 & 6.102 & 2.498 \\
11 & 10.14 & 5.148 & 11 & 2.496 & 11 & 42.57 & 19.58 \\
13 & 8.916 & 4.469 & 13 & 1.946 & 13 & 37.41 & 18.51 \\
15 & 7.776 & 3.890 & 15 & 1.726 & 15 & 27.66 & 13.89 \\
17 & 6.883 & 3.424 & 17 & 1.439 & 17 & 20.00 & 9.930 \\
19 & 6.132 & 3.004 & 19 & 1.199 & 19 & 14.41 & 7.187 \\
21 & 5.436 & 2.650 & 21 & 1.073 & 21 & 10.53 & 5.190 \\
23 & 4.825 & 2.296 & 23 & 0.9512 & 23 & 8.007 & 3.863 \\
25 & 4.357 & 2.098 & 25 & 0.8151 & 25 & 6.028 & 2.911 \\
27 & 3.959 & 1.875 & 27 & 0.7535 & 27 & 4.583 & 2.202 \\
29 & 3.545 & 1.666 & 29 & 0.6718 & 29 & 3.577 & 1.721 \\
31 & 3.208 & 1.526 & 31 & 0.5796 & 31 & 2.879 & 1.342 \\
33 & 2.950 & 1.367 & 33 & 0.5276 & 33 & 2.248 & 1.078 \\
35 & 2.683 & 1.207 & 35 & 0.5491 & 35 & 1.813 & 0.8730 \\
37 & 2.468 & 1.131 & 37 & 0.4692 & 37 & 1.507 & 0.7100 \\
39 & 2.255 & 1.034 & 39 & 0.4334 & 39 & 1.261 & 0.5682 \\ \hline
\end{tabular}
\caption{}[The rapidity-integrated $p_T^2$ distribution is given for
single $b$ quarks(NLO and Born) and the $p_T^2$ distribution in the range
$|y|<1$ is given for $b \overline{b}$ pair production (NLO only).
The
tabulated results have not been corrected for the rapidity bin width.
The rapidity-integrated pair mass distribution is also given.  All
distributions
are at $\sqrt{s}=5.5$ TeV and calculated with MRS D-$^\prime$ parton
distributions.  Note the 2 GeV bin width for the distributions.]
\end{table}
\newpage

\begin{table}
\begin{tabular}{|c|c|c||c|c||c|c|c|}
\multicolumn{8}{c}{{\bf $b \overline{b}$ Production $\sqrt{s}=14$ TeV}}
\\ \hline
\multicolumn{3}{c||}{$d\sigma_b/dp_T^2$ ($\mu$b/2 GeV$^2$)} &
\multicolumn{2}{c||}{$d\sigma_{b \overline{b}}/dp_T^2dy$ ($\mu$b/2 GeV$^2$)} &
\multicolumn{3}{c}{$d\sigma_{b \overline{b}}/dM$ ($\mu$b/2 GeV)} \\ \hline
$p_T^2$ (GeV$^2$) & NLO & LO & $p_T^2$ (GeV$^2$) & NLO & $M$ (GeV) & NLO & LO
\\ \hline
1  & 68.43 & 32.54 & 1 & -13.36 & & & \\
3  & 56.73 & 28.24 & 3 & 34.99 & & & \\
5  & 47.74 & 24.25 & 5 & 17.94 & & & \\
7  & 41.32 & 20.92 & 7 & 11.83 & & & \\
9  & 35.45 & 18.10 & 9 & 8.519 & 9 & 17.57 & 6.876 \\
11 & 30.61 & 15.55 & 11 & 6.833 & 11 & 124.0 & 55.90 \\
13 & 27.07 & 13.60 & 13 & 5.537 & 13 & 112.4 & 54.74 \\
15 & 23.97 & 11.93 & 15 & 4.665 & 15 & 85.11 & 42.17 \\
17 & 21.22 & 10.41 & 17 & 3.813 & 17 & 62.92 & 30.97 \\
19 & 18.86 & 9.192 & 19 & 3.392 & 19 & 46.41 & 22.58 \\
21 & 16.84 & 8.225 & 21 & 3.125 & 21 & 34.27 & 16.62 \\
23 & 15.20 & 7.227 & 23 & 2.618 & 23 & 26.12 & 12.44 \\
25 & 13.71 & 6.477 & 25 & 2.328 & 25 & 19.89 & 9.457 \\
27 & 12.61 & 5.878 & 27 & 2.112 & 27 & 15.51 & 7.304 \\
29 & 11.20 & 5.215 & 29 & 1.772 & 29 & 11.93 & 5.673 \\
31 & 10.43 & 4.710 & 31 & 1.811 & 31 & 9.610 & 4.538 \\
33 & 9.520 & 4.368 & 33 & 1.588 & 33 & 7.908 & 3.587 \\
35 & 8.651 & 3.962 & 35 & 1.409 & 35 & 6.267 & 2.966 \\
37 & 7.795 & 3.492 & 37 & 1.349 & 37 & 5.132 & 2.402 \\
39 & 7.272 & 3.245 & 39 & 1.279 & 39 & 4.323 & 2.017 \\ \hline
\end{tabular}
\caption{}[The rapidity-integrated $p_T^2$ distribution is given for
single $b$ quarks(NLO and Born) and the $p_T^2$ distribution in the range
$|y|<1$ is given for $b \overline{b}$ pair production (NLO only).
The
tabulated results have not been corrected for the rapidity bin width.
The rapidity-integrated pair mass distribution is also given.  All
distributions
are at $\sqrt{s}=14$ TeV and calculated with MRS D-$^\prime$ parton
distributions.  Note the 2 GeV bin width for the distributions.]
\end{table}
\newpage
\clearpage

\begin{center}
{\bf Figure Captions}
\end{center}
\vspace{0.2in}
\noindent 1.  (a) Gluon distributions from GRV HO (solid), MRS D0$^\prime$
(dashed), MRS D-$^\prime$ (dot-dashed) at $Q=2.4$ GeV and GRV HO (dotted) at
$Q= 1.2$ GeV. (b) The running of the coupling constant with scale.

\vspace{0.2in}
\noindent 2.  Investigation of uncertainties in the total cross section as
a function of scale.  Variation of the $c\overline{c}$
production cross sections with
scale at (a) RHIC and (b) LHC. Variation of the $b\overline{b}$ production
cross sections with scale at (c) RHIC and (d) LHC. Variation of the
$c\overline{c}$ production cross sections at $\sqrt{s}$ at 200 GeV with
$\mu_R$ at fixed $\mu_F$ (e) and with $\mu_F$ at fixed $\mu_R$ (b).  In each
case, the circles represent the NLO calculation, the crosses, the Born
calculation.

\vspace{0.2in}
\noindent 3.  Total charm production cross sections from $pp$ and $pA$
measurements \cite{Reu,AMM,E6531,NA27,NA321} compared to calculations.  The
curves are:  MRS D-$^\prime$ $m_c=1.2$ GeV, $\mu = 2m_c$ (solid); MRS
D0$^\prime$ $m_c=1.2$ GeV, $\mu = 2m_c$ (dashed); GRV HO $m_c = 1.3$ GeV,
$\mu = m_c$ (dot-dashed); GRV HO $m_c = 1.5$ GeV, $\mu = m_c$ (dotted).

\vspace{0.2in}
\noindent 4.  (a) Total charm production cross sections from $\pi^- p$
measurements \cite{Reu,E515,NA322,E769,E6532} compared to calculations. The
curves are:  GRV HO $m_c = 1.3$ GeV,
$\mu = m_c$ on a nucleon (solid) and proton target (dashed);
MRS D-$^\prime$ $m_c=1.2$ GeV, $\mu = 2m_c$ with SMRS P2 (dot-dashed) and SMRS
P1 (dotted) on a proton target.  (b) The $b \overline{b}$ production cross
section from $\pi^- p$ interactions \cite{NA10,E6533,WA78,E672}.  The
calculations use $m_b = 4.75$ GeV and $\mu = m_b$.  The curves use GRV HO
(solid) and MRS D-$^\prime$ with SMRS P1 (dashed).

\vspace{0.2in}
\noindent 5.  Comparison with $D$ meson (a) $p_T^2$ and (b) $x_F$ distributions
at 800 GeV \cite{AMM}.  The NLO calculations are with MRS D-$^\prime$ (solid)
and GRV HO (dashed) parton distributions.
\vspace{0.2in}

\noindent 6. Comparison with $D\overline{D}$ production for (a) $p_T^2$ and (b)
$M$ and (c) $\phi$ at 800 GeV \cite{Emu}.  The NLO calculations are with
MRS D-$^\prime$ (solid) and GRV HO (dashed) parton distributions.
\vspace{0.2in}

\noindent 7.  Comparison with $b$ quark production cross sections at
(a) UA1 \cite{UA1} and (b) CDF \cite{CDF2}.  The NLO calculations are with
MRS D-$^\prime$ (solid) and GRV HO (dashed) parton distributions.
\vspace{0.2in}

\noindent 8.  Predictions for $c$ and $c \overline{c}$ production at $\sqrt{s}
=
200$ GeV with MRS D-$^\prime$ distributions.  The $c$ quark $p_T$ distributions
at NLO (solid)  are shown in (a) and the rapidity distributions at LO (dashed)
and NLO (solid) are shown in (b). The $c \overline{c}$ pair distributions are
shown in (c)-(f).  The LO (dashed) distributions are shown only for mass and
rapidity.  Additionally, the $p_T$ and $p_{T_p}$ distributions are shown with a
central cut in rapidity.  (The rapidity bin widths are removed.)  The
corresponding distributions from HIJING are also shown, again with the rapidity
bin width divided out.
\vspace{0.2in}

\noindent 9.  Predictions for $c$ and $c \overline{c}$ production at $\sqrt{s}
= 200$ GeV with GRV HO distributions. The corresponding distributions
from HIJING are also shown.
\vspace{0.2in}

\noindent 10.  Predictions for $c$ and $c \overline{c}$ production at $\sqrt{s}
= 500$ GeV with MRS D-$^\prime$ distributions.
\vspace{0.2in}

\noindent 11.  Predictions for $c$ and $c \overline{c}$ production at $\sqrt{s}
= 500$ GeV with GRV HO distributions.
\vspace{0.2in}

\noindent 12.  Predictions for $c$ and $c \overline{c}$ production at $\sqrt{s}
= 5.5$ TeV with MRS D-$^\prime$ distributions.  The corresponding
distributions from HIJING are also shown.
\vspace{0.2in}

\noindent 13.  Predictions for $c$ and $c \overline{c}$ production at $\sqrt{s}
= 5.5$ TeV with GRV HO distributions.  The corresponding distributions from
HIJING are also shown.
\vspace{0.2in}

\noindent 14.  Predictions for $c$ and $c \overline{c}$ production at $\sqrt{s}
= 14$ TeV with MRS D-$^\prime$ distributions.
\vspace{0.2in}

\noindent 15.  Predictions for $c$ and $c \overline{c}$ production at $\sqrt{s}
= 14$ TeV with GRV HO distributions.
\vspace{0.2in}

\noindent 16.  Predictions for $b$ and $b \overline{b}$ production at $\sqrt{s}
= 200$ GeV with MRS D-$^\prime$ distributions.
\vspace{0.2in}

\noindent 17.  Predictions for $b$ and $b \overline{b}$ production at $\sqrt{s}
= 200$ GeV with GRV HO distributions.
\vspace{0.2in}

\noindent 18.  Predictions for $b$ and $b \overline{b}$ production at $\sqrt{s}
= 500$ GeV with MRS D-$^\prime$ distributions.
\vspace{0.2in}

\noindent 19.  Predictions for $b$ and $b \overline{b}$ production at $\sqrt{s}
= 500$ GeV with GRV HO distributions.
\vspace{0.2in}

\noindent 20.  Predictions for $b$ and $b \overline{b}$ production at $\sqrt{s}
= 5.5$ TeV with MRS D-$^\prime$ distributions.
\vspace{0.2in}

\noindent 21.  Predictions for $b$ and $b \overline{b}$ production at $\sqrt{s}
= 5.5$ TeV with GRV HO distributions.
\vspace{0.2in}

\noindent 22.  Predictions for $b$ and $b \overline{b}$ production at $\sqrt{s}
= 14$ TeV with MRS D-$^\prime$ distributions.
\vspace{0.2in}

\noindent 23.  Predictions for $b$ and $b \overline{b}$ production at $\sqrt{s}
= 14$ TeV with GRV HO distributions.
\vspace{0.2in}

\noindent 24.  Dilepton (a) mass, (b) rapidity, and (c) $p_T$ distributions
at $\sqrt{s}$ = 200 GeV
from $c \overline{c}$ decays calculated using MRS D-$^\prime$ distributions are
shown.
\vspace{0.2in}

\noindent 25.  Dilepton distributions at $\sqrt{s}$ = 200 GeV
from $b \overline{b}$ decays.
\vspace{0.2in}

\noindent 26.  Dilepton distributions at $\sqrt{s}$ = 5.5 TeV
from correlated $c \overline{c}$ decays.
\vspace{0.2in}

\noindent 27.  Dilepton distributions at $\sqrt{s}$ = 5.5 TeV
from uncorrelated $c \overline{c}$ decays.
\vspace{0.2in}

\noindent 28.  Dilepton distributions at $\sqrt{s}$ = 5.5 TeV
from $b \overline{b}$ decays.

\end{document}